# Parsing Data Formats of the Inputs and Outputs of Geographic Models with Code Analysis

Model web services provide an approach for implementing and facilitating the sharing of geographic models.  The description and acquisition of inputs and outputs (I/O) of geographic models is a key issue in constructing and using model web services.  These approaches for describing and acquiring the data formats of the I/O of geographic models can be classified into two categories, i.e., intermediate-data-format-based and native-data-format-based.  However, these two categories mainly consider the description of the I/O of geographical models but relatively pay little attention to the acquisition.  To address this issue, this paper proposes an approach for automatically parsing data formats of the I/O utilizing the relationship between the I/O and source codes.  This proposed approach can utilize such a strict and coupling relationship and the expression form of the data formats in the source codes to retrospectively derive the I/O data format and automatically generate data format documentation.  The feasibility of the proposed approach has been verified via a geographical model coded in the FORTRAN language, which shows that it significantly improves the efficiency of writing data format specifications and promotes sharing geographic models as model web services.

This paper was finished on 10 Aug 2018. A new paper will be submitted to Environmental Modelling & Software

Xinghua Cheng[a,d], Di Hu[a,b,c]*, Handong He[e], Guonian Lv[a,b,c] A-Xing Zhu[b,f]

[a] *School of Geography, Nanjing Normal University, Nanjing, China;*

[b] *Jiangsu Center for Collaborative Innovation in Geographical Information Resource Development and Application, Nanjing, China;*

[c] *Key Laboratory of Virtual Geographic Environment (Nanjing Normal University), Ministry of Education, Nanjing, China;*

[d] *Department of Land Surveying and Geo-Informatics, The Hong Kong Polytechnic University, Hong Kong, China;*

[e] *School of Resources and Environment, Anhui Agricultural University, Hefei, China;*

[f] *Department of Geography, University of Wisconsin-Madison, Madison, USA*

Corresponding author: Di Hu, hud316@njnu.edu.cn, School of Geography, NanJing Normal University, No.1, Wenyuan Road, Xianlin University District，Nanjing, China.

This paper was finished on 10 Aug 2018. A new paper will be submitted to Environmental Modelling & Software

**1. Introduction**

Sharing geographic models is vital for geography research. It enables geographers to reduce the duplication of geographic model programming and implementation, instead to focus on their research issues (Crosier et al. 2003, Granell et al. 2010, Tang et al. 2006, Feng et al. 2006). Model web service is the most used approach for geographic models sharing (Zhai et al. 2016, Yang et al. 2015, Geller & Melton 2008, Nativi et al. 2013, Castronova et al. 2013, Goodchild et al. 2007, Chen et al. 2014, Wen et al. 2017). To publish a geographic model as a model web service, data formats of the inputs and outputs (I/O) of this geographic model should be clearly described and shared (Hu et al. 2015). When using a model web service, model users should clearly know the data formats of the I/O. According to the data formats, model users need to know how to prepare the input data and to understand the output data of the model web service. Therefore, describing and acquiring the data formats of the I/O of geographic models is a key issue in constructing and using model web service.

According to the data formats adopted by the model web service, current approaches describing and acquiring the data formats of the I/O of geographic models can be mainly classified into two types: intermediate data format-based approach and native-data-format-based approach. The basic idea of the former is that the model web service uses intermediate data formats, e.g. Extensive Markup Language (XML), and use XML schema to describe the intermediate data formats (W3C 2019, Microsoft 2019). The model web services implemented by web services containing Web Processing Service (WPS) (Open Geospatial Consortium 2007) belong to this type. Specifically, web services describe the intermediate data formats using basic programming data types, e.g. integer, string, float and double, as well as their efficacious combinations. This means that it can describe the data types and structures of some simple data and complex data. Regarding WPS, which aims to the GIS data, utilizes some specific data types, e.g. BoundingBox defined as a minimum bounding rectangle in geographic coordinates, and add file types, e.g. Binary files, XML files, and images. This indicates that WPS mainly concerns the issues of the data formats of files but does not describe the data formats of files in details. In fact, the data formats of files should be described from aspects of data type, structure and layout. Some details such as the separators, location of data items is crucial important for data formats of files. Since most I/O of geographic models are files, the data formats of the I/O of geographic models cannot perfectly be described by intermediate-data-format-based approaches.

The basic idea of native-data-format-based approach is that the model web services use the native data formats of I/O of geographic models, describing the native data formats clearly in a universal way. In this sense, a data format description language was proposed, namely Data Format Markup Language (DFML) (Hu et al. 2015). DFML not only provide data types, but also separators, location and their combination. These are necessary for describing the data format of a file, because a file consists of many data items and separators. The location and combination of data items and separators determines the data format of a file. Since most geographical model I/O are files, describing the data formats of geographic models can be well done using DFML.

These two approaches mentioned above only involves the description of I/O of geographic models (W3C 2019). With regards to the acquisition of the data formats of the I/O of geographic models, both pay little attention to it. Moreover, to the best of our knowledge, no study has examined how to acquire the data formats of I/O of geographic models. Hence, in all probability, these two types of approaches assume that the data formats of I/O of geographic models are acquired in a manual way. Manual acquisition of data formats of geographic models is not only tedious but also error prone. Manual acquisition is extremely difficult for some legacy models because of the lack of documentation on this I/O formats. Thus, for the information acquisition of models on I/O formats requires manual crawling through the source codes of the model programs, which is prohibitively difficult.

In this paper, to address the issue of acquiring the I/O of geographic models, an approach is proposed to automatically parse the data format of the I/O through the source codes. The remainder of this article is organized as follows: Section 2 analyses the relationship between the source code and the data format of the I/O of geographic models. Section 3 presents the design of the proposed approach that parsed the source code of geographic models and extracted and structurally organized the data format information of the I/O files. Section 4 describes the implementation of the approach taking geographic models written in FORTRAN as an example. Conclusions are made in Section 5.

**2. Relationship between source code and data format**

When a geographical model developer writes program code, the data are usually read by following the input data format. Using various calculation processes, the data will be written into an output file following the designated output format (De et al. 2000). This type of strict correspondence between the geographical model program code and the data format results in a high degree of coupling between the two, which produces strict requirements on the data format of the geographical model. However, we can utilize this strict and coupling relationship as well as the expression form of the data format in the source code to retrospectively derive the input and output data format and automatically generate data format documentation.

At present, a uniform understanding and definition of the data format has not been implemented (Zhou et al. 1996, Li et al. 2012). According to Horak (2007), a data format is defined as "a critical part of a communications protocol that enables the receiving device to logically determine what is to be done with the data and how to go about doing it". A data format consists of three parts: a header, text, and trailer. The text stores data content, and the header and trailer contribute to the successful transfer of the data content (Horak 2007). A large number of domain-specific data description methods have emerged in recent years (Levy et al. 1996, Mandelbaum et al. 2007, Fernández et al .2008). In the present case, the data format is defined by three parts, i.e. the data type, layout, and structure. To clearly describe the relationship between the geographical model and the data format, all these parts are considered in this paper. Using a combination of data types, layout, and structure, the data type of an item at a certain location can be unambiguously determined along with the specific expression form of the item. Thus, the type of the data format can be unambiguously determined (Fisher et al. 2006). For the three aspects considered, the data type determines the data type of each item in a global data set, the data layout indicates the location of a specific item and its expression form, and the data structure represents the contents of a global data set and the relationships between the items

Various input and output parameters of computers are utilized by read and write operations in high-level programming languages. A read and write statement is responsible for informing the computer of the data that should be input and output, as well as the input and output format and origin and destination of the input and output. On an operation systems level, the computer uses files to manage external devices and various types of information. All inputs and outputs will eventually be read from and written to files. For example, when the input device is a keyboard, this could be considered reading a file from a standard input device; when the output device is a control panel, this could be considered writing a file to a standard output device.

All inputs and outputs of a geographical model are usually completed simultaneously by a series of read and write operation statements, constant definition statements, format statements, and flow control statements. Variables and constants are the direct representations of data input and output; the data type that is usually indicated in the variable and constant definition statement corresponds to the data type information in the data format. Read and write operation statements contain the format information of data; the concrete manifestations include format descriptors and format description statements. These statements define the input and output control format of each type of data and correspond to the layout information of the data format. The data format structure is a combination and repetition of a specific data type and the layout information, which results in a specific read and write operation sequence in the source codes. Read and write operation statements are imbedded in specific flow control statements; they form specific sequences of read and write operations and correspond to the structure information in the data format. For example, the reading and writing operation statements imbedded in a looping structure form a repeated read and write operation sequence with a specific regularity, corresponding to the rule structure of the data format. Figure 1 shows the corresponding relationship between the main syntax structure and the data format in the source code.

This paper was finished on 10 Aug 2018. A new paper will be submitted to Environmental Modelling & Software

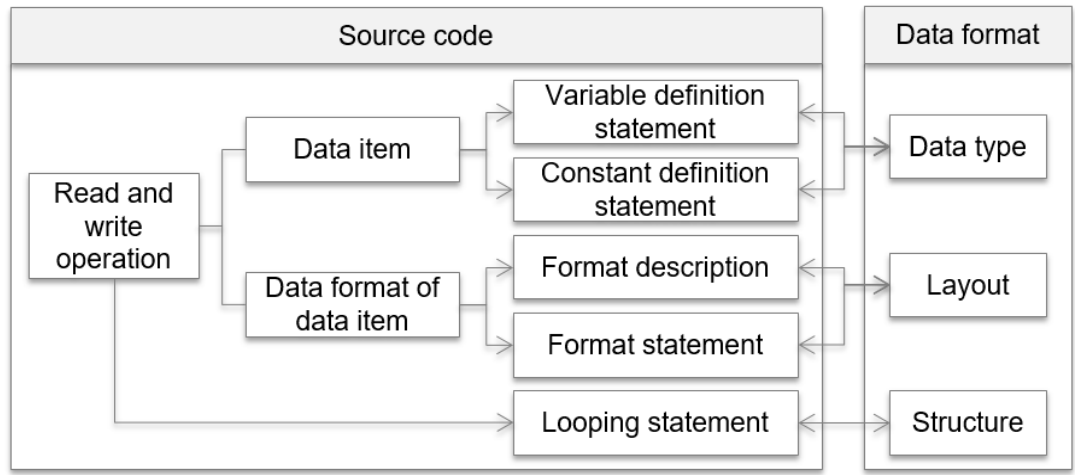

**Figure** 1. Corresponding relationship between the source code and the data format

In the following, a segment of a geographical model written in FORTRAN language is used as an example (Wang 2006) to specifically analyze the relationship between the source code and the data format. The parts of the source code of the module are as follows:

```
1    PARAMETER (N=136)
2    DOUBLE PRECISION D(N,N)
3    DOUBLE PRECISION BEMP(N),POP(N),SEMP(N)
4    INTEGER I,J,OZONE(N),DZONE(N)
5    OPEN (2, FILE='ODTIME.PRN', STATUS='OLD')
6        DO 10 I=1,N
7            DO 10 J=1,N
8                READ (2,*) OZONE(I), DZONE(J), D(I,J)
9    10   CONTINUE
10       CLOSE(2)
11   OPEN(12,FILE='Basic.TXT')
12       DO 500 I=1,N
13           WRITE(12,501) I,OZONE(I),BEMP(I),POP(I),SEMP(I)
```

The first line declares and assigns a value to a constant N (default type, integer); the second line declares a double-precision, floating-point, one-dimensional array

This paper was finished on 10 Aug 2018. A new paper will be submitted to Environmental Modelling & Software variable; the third line declares four double-precision, one-dimensional arrays and one double-precision, two-dimensional array; and the fourth line declares three integer variables and three integer one-dimensional arrays.

From line 5 to line 10, the code reads data from the file ODTIME.PRN. The READ statement in line 8 reads a line of data from the file, and the read order is as follows: integer variable, integer variable, and double-precision, floating point variable. Every line of the file contains two integer type data points and a real type data point, and FORTRAN default separators are present between the data. Lines 6-9 use a double loop to read data from the file for a total of N*N times (136*136 times); thus, an ODTIME.PRN file with 18,496 lines satisfies the module.

Lines 11-16 write the data into the file Basic.TXT. Line 13 writes an integer variable, two additional integer variables, and three double precision floating point variables using the WRITE statement, and the output format is specified by line label 501. Lines 12-15 write the data into the file using a double loop, for a total of N times (136 times); thus, a Basic.TXT file with136 lines satisfies the module. The "1X,2(1x,i4),3(1X,f12.6)" of the format descriptor in the FORMAT statement in line 14 that represents the data format layout is "space, space, integer with a width of four, space, integer with a width of four, space, real number with a width of 12 and six digits after the decimal point, space, real number with a width of 12 and six digits after a decimal point, space, and real number with a width of 12 and six digits after the decimal point", as shown in Figure 2.

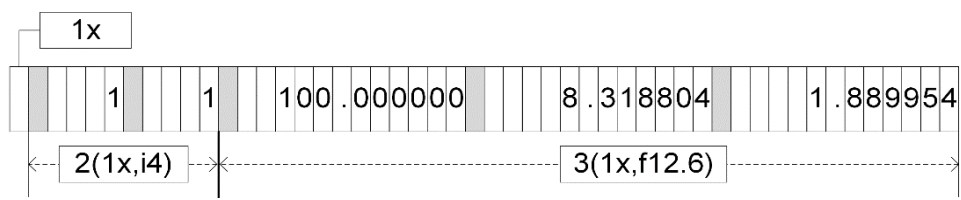

**Figure** 2. An example of the data format layout

## 3. Data format parsing of the input and output files of the geographical model based on code analysis

### *3.1. Basic principles and technical processes*

A word is the smallest syntax unit that has real meaning in high-level programming languages, and each word is composed of characters. Every high-level programming language defines a character set. Lexical analysis is used to identify each of these words input from the source code character sequence. Every high-level programming language has a set of rules, called syntax rules or grammar. Based on the syntax, words can form syntax units, such as statements, processes, and procedures. Syntax analysis is a process used to determine whether the input sequence can form syntactically correct statements and procedures through syntax decomposition (Qin 1995, Berzal 2015).

  A program used for lexical analysis is called a lexical analyzer, and a program or a function used for syntax analysis is called a syntax analyzer. Tools such as Lex and Yacc can be used in code parsing (Huang et al. 2011, Meyer 2007), text input processing (Zhang et al. 1998), JSON format parsing [Honda & Kuramitsu, 2015]; and file parsing (Wang& Lin 2010, Mutalik et al. 2001, Fox 1991); they can perform relevant semantic operations while analyzing program codes to achieve certain application goals.

  Based on the correspondence of the geographical model source code and data format and by referencing the idea of lexical and syntax analysis, this paper proposes an automatic parsing method for input and output file formats for geographical models. This method includes three stages of data format parsing that are oriented toward the specific construction of a lexical analyzer, construction of a syntax analyzer, and source code parsing of the geographical model. The key to the method lies in the construction of specific lexical and syntax analyzers for data format parsing. Therefore, it is only necessary to call the corresponding lexical and syntax analyzer based on the source code of the specific geographical model, for example, at the source code parsing stage to automatically generate input and output data format documentation files. The technical procedure is shown in Figure 3.

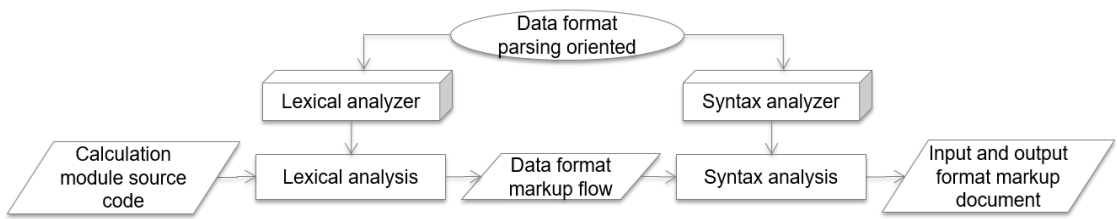

**Figure** 3. Technical procedure of the method

A data format-parsing-oriented specific lexical analyzer follows the general flow of lexical analyzer construction; the key is to identify a particular label based on the expressions and features of the geographical model data format information in the program language and pass the label to a syntax analyzer. A data format-parsing-oriented specific syntax analyzer also follows the general flow of syntax analyzer construction, and the key is to identify a specific syntax structure from the syntax features of the data format label in the programming language and perform corresponding semantic processing by parsing out data format information and generating a format documentation file after structural organization.

*3.2. Construction method for a lexical analyzer*

The general construction flow of a lexical analyzer is shown in Figure 4. The difference in this method is that the classification and definition of the labels are performed for the code elements in the geographical model source code and they are related to the data format.

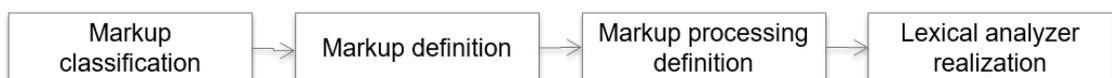

**Figure 4.** Lexical analyzer construction procedure

(1) Label classification. The character set in the programming language used in the geographical model is classified based on the relationship between the geographical model source code and the data format with respect to the identifiers, integer constants, real constants, data type keywords, control structure keywords, file operation keywords, read and write operation keywords, data format descriptors, control format descriptors, and other characters.

(2) Label definition. According to the commonly used syntax rules of the programming language used in the geographical model, regular expressions are used to define the composition structure of the labels. Identifiers, integer constants, real constants, data format descriptors, and control format descriptors are defined as types of labels; each keyword in the data type keywords, control structure keywords, file operation keywords, and read and write operation keywords is defined as a type of label.

(3) Label processing definition. For the labels defined in (2), the value and classification of the labels are stored and passed to the syntax analyser.

(4) Program codes are written, or tools such as Lex are used to generate a lexical analyser.

### *3.3. Construction method for the syntax analyser*

The general construction flow of a syntax analyzer is shown in Figure 5. The proposed method only needs to identify the syntax structure corresponding to the data format and to add semantic processing actions to extract data format information; semantic processing is the key procedure.

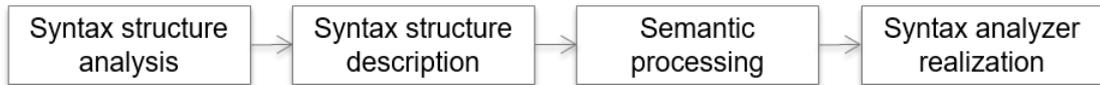

**Figure** 5. General flow of syntax analyzer construction

- Syntax structure analysis. In geographical model source codes, code elements that correspond to the module data format are embedded in syntax structures such as the program structure, constant and variable declarations, file operation statements, read and write operation statements, format statements and looping structures. Syntax structure analysis is performed to analyze the features and specifications of these syntax structures and to analyze the logical relationship among data format-related code elements in these syntax structures.

- Syntax structure description. According to the syntax rules of the programming language, a Backus-Naur Form (BNF) paradigm is used to describe the above syntax structure.

- Semantic processing. Semantic processing was originally performed by lexical and syntax analysis during the compilation process. Currently, semantic processing commonly uses the syntax-directed translation technique. The corresponding semantic processing is performed at the same time as syntax analysis. Corresponding semantic processing is added during the syntax description, as shown in Figure 6.

The process information is extracted during the syntax description of the program structure and is added into a process table to determine the possible existence of variables and constants with same names. The variable/constant information is extracted during the description of variable/constant declaration syntax and is added to a variable/constant table for subsequent determination of the data type of input and output data. To extract the name and path of the opened file from the syntax description of the open file; a data format documentation file with the same name is created in the specified directory and open the file. To close the current data format documentation file, the syntax description of closing the file is similarly used. In the syntax description of the looping structure, the number of loops at the start of the looping structure is counted, and the loop counter is set. If the number of loops is a variable, then a self-defined default looping number is set; the loop counter is zeroed at the end of the looping structure. The current read/write operation device is extracted from the read/write statement syntax, the operating file name is parsed, and the current data format documentation file is created. The read/write operation example is extracted, and the variables and constants of the read operation example table in the variable table and the constant table are searched. The data format description is extracted, and the layout information is parsed. The loop count is checked; if it is not zero, then the extracted data type and layout information is grouped, and a new iteration number is set. The data format information is written into the current data format documentation file.

This paper was finished on 10 Aug 2018. A new paper will be submitted to Environmental Modelling & Software

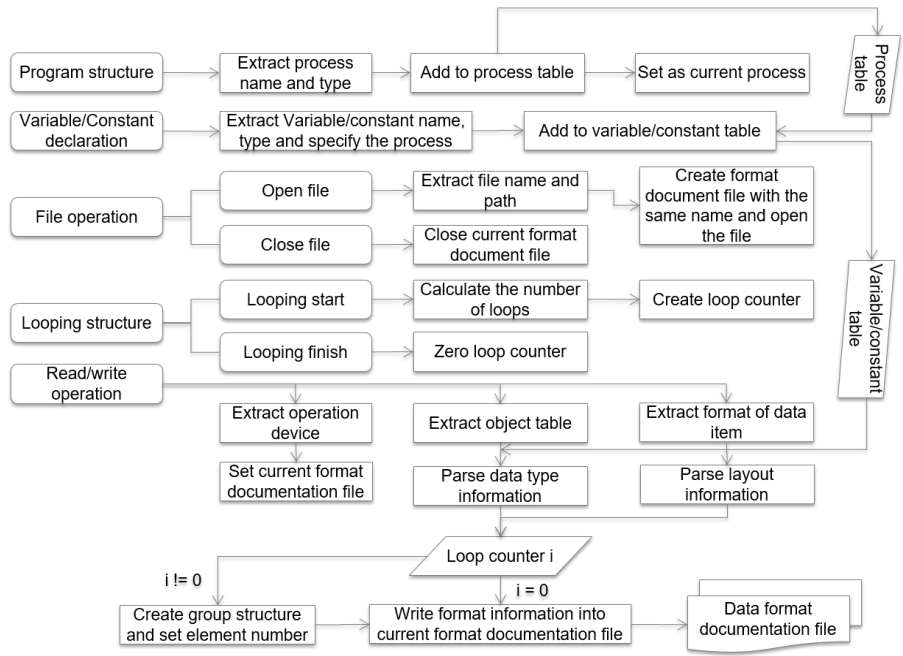

**Figure** 6. Semantic process of the data format syntax analyzer

- Program code is written, or tools such as Yacc are used to automatically generate a syntax analyser.

## 4. Program Implementation

This paper uses the FORTRAN language as an example, and C# supported GPLEX and GPPG tools are used to automatically generate lexical and syntax analyzers. Initial data format-parsing lexical and syntax analyzers are created. The semantic processing is performed by a standalone semantic processing module written in C# and is used for data format parsing, organization and output. It is provided to the syntax analyzer in a dynamic link library (DLL) form. A main control program is written to read the geographical model source code, create and call lexical and syntax analyzer examples, and generate the data format documentation file. The process flow is shown in Figure 7.

This paper was finished on 10 Aug 2018. A new paper will be submitted to Environmental Modelling & Software

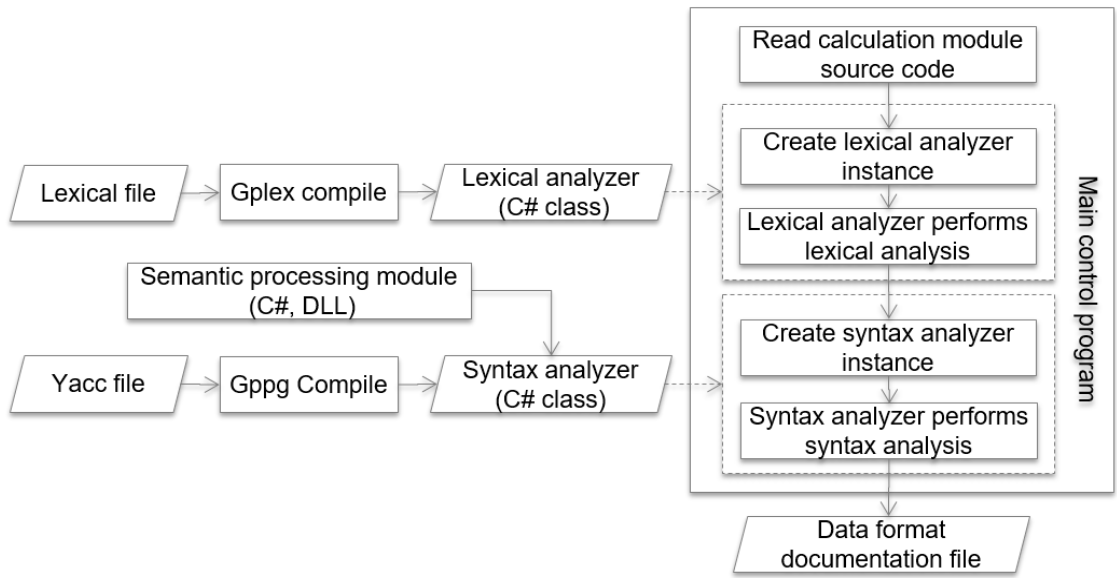

**Figure** 7. Automatic data format parsing process

The key to using the GPLEX tool to auto-generate a lexical analyzer is writing a Lex lexical file (Ding & Wang 1989, Gough 2014a). The data format label is defined according to the Lex lexical file syntax. Part of the Lex lexical file is shown, including the lexical definition and the process of the main labels.

```
real_const   [0-9]+\.[0-9]+            {integer}   {yylval.strValue=yytext;   return (int)Tokens.INTEGER;}
int_const    [0-9]+                    { real }    {yylval.strValue=yytext;   return (int)Tokens.REAL;}
identifier   [a-zA-Z_][a-zA-Z0-9_]*    { character } {yylval.strValue=yytext; return (int)Tokens.CHARACTER;}
……                                     ……
integer   integer                      { program } {yylval.strValue=yytext;   return (int)Tokens.PROGRAM;}
real      real                         { end }     {yylval.strValue=yytext;   return (int)Tokens.END;}
character character                     ……
program   program                      { open }    {yylval.strValue=yytext;   return (int)Tokens.OPEN;}
end       end                          { close }   {yylval.strValue=yytext;   return (int)Tokens.CLOSE;}
open      open                         { read }    {yylval.strValue=yytext;   return (int)Tokens.READ;}
                                       { write }   {yylval.strValue=yytext;   return (int)Tokens.WRITE;}
                                       ……
```

yytext is an attribute of the label value represented by the lexical analyzer; yylvalue is an attribute of the label value represented by the syntax analyzer, and Tokens denotes that the enumeration type will be defined in the Yacc syntax file. Using the GPLEX tool to compile the Lex lexical file, a lexical analyzer is generated, which includes a C# file containing the LexFortran type.

The key to using the GPLEX tool to auto-generate a syntax analyzer is writing a Yacc syntax file (Ding, Y.J.; Wang, N.B et al. 1989, Gough 2014b). The syntax structure is described according to syntax of the Yacc syntax file, and functionality of the semantic processing module DLL is necessary to perform the corresponding semantic process. Part of the Yacc syntax file is shown, including the syntax structure description of the variable declaration statement, the read and write operation statement, and their semantic processing.

```
declaration_part:type_spec IDENTIFIER
     { variableList.Add(new Variable($2,$1.strValue,currentProcess.Name)); }
;
type_spec: INTEGER                    { $$.strValue = $1; }
        | REAL                        { $$.strValue = $1; }
        | CHARACTER                   { $$.strValue = $1; }
        ;
```

```
read_stmt: READ IDENTIFIER

       |READ ASTERISK COMMA IDENTIFIER

       |READ LP INT_CONST COMMA ASTERISK RP IDENTIFIER

       { ···   Variable myVar = variableList.Find(delegate(Variable v)        { return v.Name == $7; });
...

        if ( loopFlag != 0)

       {  sw.WriteLine("<group number={0}", loopFlag);

         sw.WriteLine("<{0} format='{1}'> </{2}>", myVar.Type, myVar.DefaultFormat,

         myVar.Type);

             sw.WriteLine("</group>");

          }

                else

       { sw.WriteLine("<{0} format='{1}'> </{2}>", myVar.Type, myVar.DefaultFormat,

    myVar.Type); }

         }
```

Using the GPPG tool to compile the Yacc syntax file, a syntax analyzer is generated, which includes a C# file containing the LexFortran type.

C# language is used to write the main control program. First, the geographical model source code is read, and then the lexical analysis code FortranLexer and the syntax analysis code FortranParser are called in order. Next, the geographical model source code is parsed, and eventually the geographical model input and output format documentation files are generated.

## 5. Conclusions

As manually writing input and output format documentation files for a geographical model is a tedious process, this paper proposes a code analysis based on data format parsing to generate these files. This method utilizes the relationship between the geographical model source code and the input and output file formats, constructs data format-parsing-oriented specific lexical and syntax analyzers to parse the geographical model source code and generates input and output format documentation files. A geographical model written in FORTRAN is used as example for experimental verification. The results indicate that this method auto-generates data format specifications for input and output format files of a geographical model. The format file is complete and clear and can satisfy the requirements for format files in the sharing and reuse process.

Using this method, data format-parsing-oriented specific lexical and syntax analyzers can generate corresponding lexical and syntax files for different programming languages. Lexical and syntax analyzers are auto generated to support different programming languages using software tools; thus, the proposed automatic format parsing method can be widely applied to geographical models written in various programming languages. In practice, the proposed method is mainly used for the automatic parsing of large amounts user self-defined data formats, which increases the writing efficiency and standardization of the formatted files and promotes the sharing and reuse of geographical models.

**Disclosure statement:** No potential conflict of interest was reported by the authors.

This paper was finished on 10 Aug 2018. A new paper will be submitted to Environmental Modelling & Software

**References**

Berzal, Fernando, Francisco J. Cortijo, Juan-Carlos Cubero, and Luis Quesada. 2015 "The modelcc model-driven parser generator." *arXiv preprint arXiv:1501.02038*.

Crosier, Scott J., Michael F. Goodchild, Linda L. Hill, and Terence R. Smith. 2003. "Developing an infrastructure for sharing environmental models." *Environment and Planning B: Planning and Design* 30: 487-501.

Castronova, Anthony M., Jonathan L. Goodall, and Mostafa M. Elag. 2013. "Models as web services using the open geospatial consortium (ogc) web processing service (wps) standard." *Environmental Modelling & Software* 41: 72-83.

Chen, Zeqiang, Hui Lin, Min Chen, Deer Liu, Ying Bao, and Yulin Ding. 2014. "A framework for sharing and integrating remote sensing and GIS models based on Web service." *The Scientific World Journal* 2014.

Ding, Y.J.; Wang, N.B. 1989. "Analysis and applications of language implementation tools, lex and yacc." *Computer Application and Software*., 5-11. (in Chinese)

De Silva, F. Nisha, and R. W. Eglese. 2000 "Integrating simulation modelling and GIS: spatial decision support systems for evacuation planning." *Journal of the Operational Research Society* 51: 423-430.

Fisher, Kathleen, Yitzhak Mandelbaum, and David Walker. 2006 "The next 700 data description languages." In *ACM Sigplan Notices* 41:2-15.

Fernández, Mary, Kathleen Fisher, J. Nathan Foster, Michael Greenberg, and Yitzhak Mandelbaum. 2008 "A generic programming toolkit for PADS/ML: First-class upgrades for third-party developers." Paper presented at *International Symposium on Practical Aspects of Declarative Languages*, Springer, Berlin, Heidelberg, pp. 133-149.

This paper was finished on 10 Aug 2018. A new paper will be submitted to Environmental Modelling & Software

Fox, Edward A., Prabhakar M. Koushik, Qi-Fan Chen, and Robert K. France. 1991 "Integrated access to a large medical literature database.".

Feng, Min, Shiqiang Zhang, and Xin Gao. 2010 "Glacier runoff models sharing service and online simulation." Paper presented at *2010 Second International Conference on Advanced Geographic Information Systems, Applications, and Services*, IEEE, 123-126.

Goodchild, Michael F., Pinde Fu, and Paul Rich. 2007. "Sharing geographic information: an assessment of the Geospatial One-Stop." *Annals of the Association of American Geographers* 97: 250-266.

Geller, Gary N., and Forrest Melton. 2008. "Looking forward: Applying an ecological model web to assess impacts of climate change." *Biodiversity* 9: 79-83.

Granell, Carlos, Laura Díaz, and Michael Gould. 2010."Service-oriented applications for environmental models: Reusable geospatial services." *Environmental Modelling & Software* 25: 182-198.

Gough J. 2014a The GPLEX Scanner Generator (Version 1.2.2).

Gough J. 2014b The GPPG Parser Generator (Version 1.5.1).

Halevy, Alon, Anand Rajaraman, and Joann Ordille. 2006 "Data integration: the teenage years." Paper presented at *Proceedings of the 32nd international conference on Very large data bases*: 9-16.

Horak, Ray. *Telecommunications and data communications handbook*. John Wiley & Sons, 2007.

Huang, P. Wang, L.P. Guan, L.W. Yao, R. 2011. "Development of an open post-processing system for parallel machines based on Lex and Yacc." *High Technology Letters* 21:303-308. (in Chinese)

Honda, Shun, and Kimio Kuramitsu. "Implementing a Small Parsing Virtual Machine on Embedded Systems." *arXiv preprint arXiv:1511.03406* (2015).

Hu, Di, Shaosong Ma, Fei Guo, Guonian Lu, and Junzhi Liu. 2015 "Describing data formats of geographical models." *Environmental Earth Sciences* 74: 7101-7115.

This paper was finished on 10 Aug 2018. A new paper will be submitted to Environmental Modelling & Software

Levy, Alon, Anand Rajaraman, and Joann Ordille.1996. *Querying heterogeneous information sources using source descriptions*. Stanford InfoLab.

Li, N.; Liang, Q.; Shi, Y.M. 2012. "The function of format information in document understanding." *Journal of Beijing Information Science and Technology University* 27, :1-7. (in Chinese)

Mutalik, Pradeep G., Aniruddha Deshpande, and Prakash M. Nadkarni. 2001 "Use of general-purpose negation detection to augment concept indexing of medical documents: a quantitative study using the UMLS." *Journal of the American Medical Informatics Association* 8: 598-609.

Mandelbaum, Yitzhak, Kathleen Fisher, David Walker, Mary Fernandez, and Artem Gleyzer. 2007 "PADS/ML: A functional data description language." In *ACM SIGPLAN Notices* 42:77-83.

Meyer-Baese, Uwe, and U. Meyer-Baese. *Digital signal processing with field programmable gate arrays*. Vol. 65. Berlin: springer, 2007.

Microsoft, https://docs.microsoft.com/en-us/previous-versions/dotnet/netframework 2.0/7h3ystb6(v=vs.80), Accessed 20 December 2019.

Nativi, Stefano, Paolo Mazzetti, and Gary N. Geller. 2013. "Environmental model access and interoperability: The GEO Model Web initiative." *Environmental Modelling & Software* 39: 214-228.

Open Geospatial Consortium, 2007. Web Processing Service (1.0.0), Retrieved from: http://www.opengeospatial.org/standards/wps.

Qin, Z. 1995. "Compilation principle and compiler construction". PhD diss., Nanjing: Southeast University Press. (in Chinese).

This paper was finished on 10 Aug 2018. A new paper will be submitted to Environmental Modelling & Software

Tang, Zhongshi, Zhe Xie, Hongrui Zhao, and Su Zhang. 2006 "Web services based GIS model sharing service." Paper presented at *IEEE International Symposium on Geoscience and Remote Sensing*, IEEE,1598-1601.

Wang, Fahui. *Quantitative methods and applications in GIS*. CRC Press, 2006.

Wang, Wen-Long, and Zhong Lin. 2010 "Optimize of FED analysis based on technologies of lexical analysis and syntax analysis." *Computer Engineering and Design* 31: 9.

Wen, Yongning, Min Chen, Songshan Yue, Peibei Zheng, Guoqiang Peng, and Guonian Lu. 2017. "A model-service deployment strategy for collaboratively sharing geo-analysis models in an open web environment." *International journal of digital earth* 10: 405-425.

W3C WSDL 2.0. https://www.w3.org/2002/ws/#documents. Accessed 28 December 2019.

W3C, Web Services Architecture. https://www.w3.org/TR/ws-arch/wsa.pdf. Accessed 20 October 2019

Yang, Yongjun, Yaqin Sun, Songnian Li, Shaoliang Zhang, Kuoyin Wang, Huping Hou, and Shishuo Xu. 2015. "A GIS-based web approach for serving land price information." *ISPRS International Journal of Geo-Information* 4: 2078-2093.

Zhou, B.K.; Cao, Z.Q.; Chen, D.P. 1996. "The design of the classes of object-oriented finite element program." *Computational structural mechanics and applications* 13:269-278. (in Chinese)

Zhai, Xi, Peng Yue, and Mingda Zhang. 2016. "A sensor web and web service-based approach for active hydrological disaster monitoring." *ISPRS International Journal of Geo-Information* 5: 171.

Zhang, W.S. 1998 "Applications of Yacc and Lex in non-compiler programs." *Journal of Liaoning University (Natural Sciences Edition)*." 25:30-36. (in Chinese)